\documentclass{article}

\usepackage{PRIMEarxiv}

\usepackage[utf8]{inputenc} 
\usepackage[T1]{fontenc}    
\usepackage{hyperref}       
\usepackage{url}            
\usepackage{booktabs}       
\usepackage{amsfonts}       
\usepackage{nicefrac}       
\usepackage{microtype}      
\usepackage{lipsum}
\usepackage{amsmath,amssymb}
\usepackage{fancyhdr}       
\usepackage{graphicx}       
\graphicspath{{media/}}     
\title{Causes and consequences of ordering and dynamic phases of confined vortex rows in superconducting nanostripes}

\author{ \href{https://orcid.org/0000-0003-4191-9243}{\includegraphics[scale=0.06]{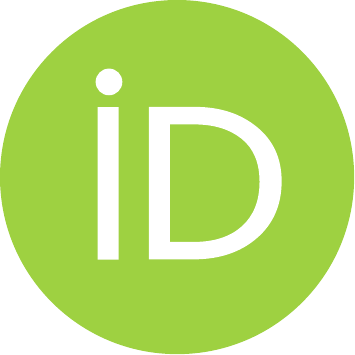}\hspace{1mm}Benjamin A. McNaughton}\\
	School of Science and Technology, Physics Division\\
	University of Camerino\\
	62032 Camerino, Italy \\
	\texttt{benjamin.mcnaughton@unicam.it} \\
	\And
	\href{https://orcid.org/0000-0001-7974-4421}{\includegraphics[scale=0.06]{orcid.pdf}\hspace{1mm}Nicola Pinto} \\
	School of Science and Technology, Physics Division\\
	University of Camerino\\
	62032 Camerino, Italy \\
	\texttt{nicola.pinto@unicam.it} \\
    \And
	\href{https://orcid.org/0000-0002-4914-4975}{\includegraphics[scale=0.06]{orcid.pdf}\hspace{1mm}Andrea Perali} \\
	School of Pharmacy, Physics Unit\\
	University of Camerino\\
	62032 Camerino, Italy \\
	\texttt{andrea.perali@unicam.it} \\
    \And
	\href{https://orcid.org/0000-0002-5431-377X}{\includegraphics[scale=0.06]{orcid.pdf}\hspace{1mm}Milorad V. Milo\v{s}evi\'{c}} \\
	Department of Physics\\
	University of Antwerp\\
	Antwerp, Belgium\\
	\texttt{milorad.milosevic@uantwerpen.be} \\
}

\begin{document}
\maketitle

\begin{abstract}
Understanding the behaviour of vortices under nanoscale confinement in superconducting circuits is of importance for development of superconducting electronics and quantum technologies. Using numerical simulations based on the Ginzburg-Landau theory for non-homogeneous superconductivity in the presence of magnetic fields, we detail how lateral confinement organises vortices in a long superconducting nanostripe, and present a phase diagram of vortex configurations as a function of the stripe width and magnetic field. We discuss why average vortex density is reduced and reveal that confinement also has profound influence on vortex dynamics in the dissipative regime under sourced electrical current, mapping out transitions between asynchronous and synchronous vortex rows crossing the nanostripe as the current is varied. Synchronous crossings are of particular interest, since they cause single-mode modulations in the voltage drop along the stripe in a high (typically GHz-to-THz) frequency range.
\end{abstract}


\section{Introduction}
Superconducting nanostripes (SN) are a fundamental component in superconducting electronics, crucial for various applications in the field of quantum technology. For example, superconducting nanostripe single-photon detectors (SNSPD) are used for quantum communication and applications in astronomy and spectroscopy \cite{hadfield2009single,natarajan2012superconducting,gol2001picosecond,dauler2007multi}. Other superconducting electronics include prototypical logic devices \cite{vlasko2017magnetic,vlasko2017manipulating,cordoba2019long}, flux qubits used in quantum computers \cite{brooke1999quantum,johnson2010scalable,strauch2003quantum}, diodes \cite{reichhardt2010jamming,wambaugh1999superconducting,daido2022intrinsic} and electromagnetic resonators \cite{samkharadze2016high,bulaevskii2006electromagnetic,dobrovolskiy2018microwave}. 
Narrow SN experience an enhancement of critical parameters \cite{perali1996gap,pinto2018dimensional,guidini2016bcs,saraiva2020multiband} due to confinement forces acting on the superconducting condensate \cite{moshchalkov1995effect,datta1997electronic,cren2009ultimate,marrocco2010strong,bean1964surface,flammia2018superconducting}. Such confinement in narrow SNs can cause large magnetoresistance oscillations \cite{berdiyorov2012magnetoresistance,anderson1964radio,cordoba2013magnetic}, where time-averaged voltage/resistance, as a function of the applied magnetic field, exhibits pronounced peaks at alternating transitions between static and dynamic vortex phases. At higher applied fields with multiple rows of vortices, or high currents, a continuous motion of vortices causes a monotonic background on which the resistance oscillations due to entries of additional vortices are superimposed \cite{berdiyorov2012magnetoresistance,berdiyorov2012large}. Commensurate effects between the SN width $w$, and the number of vortex rows $n$, have also been seen in the critical current as a function of the out-of-plane magnetic field $H$ (for fixed $w$) or $w$ (for fixed $H$) \cite{kimmel2019edge,vodolazov2013vortex}. 
Optimized operation of some of the suggested superconducting electronics may be achieved on a specific geometry of vortices.
For example, a single row of vortices was found preferable in Ref. \cite{cordoba2019long}, producing a giant non-local electrical resistance from vortices moving very far (several microns) from the local current drive. This effect appears important for a feasible long-range information transfer by vortices unaltered by the passing current.

Moving vortices however exceed in importance the bare transfer of information. For example, vortices coherently crossing SNs can produce electromagnetic radiation \cite{bulaevskii2006electromagnetic,dobrovolskiy2018microwave}, where higher radiation power is emitted when multiple vortices exit the SN simultaneously. In narrow SNs, rows of vortices can cross the SN asynchronously and synchronously \cite{berdiyorov2012magnetoresistance,vodolazov2007rearrangement}, depending on competing forces (confinement, vortex-vortex interaction, Lorentzian forces) but criteria for synchronous crossings are not yet well understood. In this respect, a study on the behaviour of vortices in SNs with small widths is important to reveal favoured geometry of vortices for the static case (no sourced current), and the relation to the dynamic case (with sourced current). Understanding how a vortex lattice is affected by the interaction with the edge confining force and other dynamic forces is important when considering SNs for applications mentioned above. Studying the dynamic dissipative states under strong confinement in the 1D-2D crossover regime, can reveal how vortices cross the SN under different conditions ($w$, $H$, current intensity). Moreover information on the possible vortex velocity under confinement \cite{jelic2016velocimetry,embon2017imaging}, can be important to both the fast information transfer and the frequency of emitted radiation by moving vortices.
Therefore, in this work we investigate how confinement in SNs affects the vortex configurations, using Ginzburg-Landau simulations \cite{milovsevic2010ginzburg}. We provide the vortex row phase diagram as a function of $H$ for a given $w$. Investigation of magnetic field dependence of the average number of vortices reveals strong confinement effects. With increasing width, reconfiguration from the vortex rows to the vortex lattice takes place, offering a criterion to define quasi 1D -to- 2D dimensional-crossover where SN effectively becomes a nanofilm in terms of the superconducting properties.
Additionally, a commensurate behaviour of the critical current, $J_{c1}(H)$, has been found when varying $H$, using a time-dependent GL approach to simulate effects of the sourced current. We show that the local minima values in $J_{c1}(H)$ (defined as the onset of vortex motion and corresponding dissipation) directly relate to row transitions shown in our vortex-row phase diagram.

Further simulations of current-voltage (I-V) characteristics in SNs have evidenced transitions among different resistive regimes (Meissner, flux-flow, flux-flow instability, phase slips, normal state). I-V curves showing similar features to our simulations for SNs have been experimentally measured only for wider structures \cite{carapella2016current, dobrovolskiy2020ultra}. 
We find that for SN with average vortex density $\lesssim$ $1/80\xi^2$ ($\xi$ being the coherence length) in a flux-flow regime, vortices cross the SN in a periodic/continuous fashion, causing modulations of the voltage drop detectable experimentally. Such a periodic flow may produce electromagnetic radiation \cite{bulaevskii2006electromagnetic}, and features characteristic power spectra \cite{hebboul1999radio} that we report by performing fast Fourier transformation of the calculated voltage drop as a function of time during vortex motion. The recorded average vortex velocity (up to 10's Km/s) is used to discuss the washboard frequencies \cite{dobrovolskiy2018microwave,bulaevskii2006electromagnetic}, in the flux-flow regime for thin SNs of niobium \cite{dobrovolskiy2020ultra}.
Providing the vortex density is sufficiently high, we that vortex row crossings transition between quasi-synchronous to synchronous, to finally asynchronous regimes. Synchronised crossings are desirable for small-band electromagnetic emitters operating in the GHz or THz range. For typical ultra thin niobium SNs \cite{pinto2018dimensional}, modulation frequencies range in the microwave regime between 10-800 GHz. Asynchronous regimes are disruptive for a coherent emission, but host a number of local dynamic vortical transitions and transformations that are of fundamental importance for advanced devices, and unattainable otherwise.

The article is organized as follows. We first introduce the theoretical framework and methods used for the numerical simulations. We then present results and discussions of all the above-listed phenomena, using both stationary and time-dependent Ginzburg-Landau approach. Main conclusions of our work are emphasized already in the results section, before being additionally commented on in the conclusions of the article.

\section{Materials and Methods}

\begin{figure}[ht]
	\centering
	{\includegraphics[width=0.5\linewidth]
		{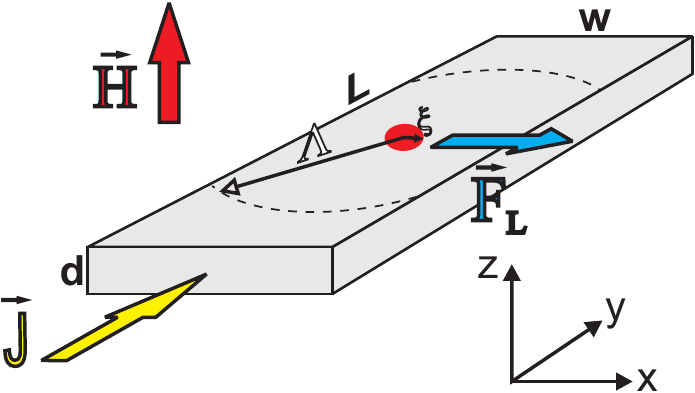}}
	\caption{Schematic illustration of a superconducting nanostripe having width $w$, length $L$ and thickness $d$, along the $x$, $y$ and $z$ directions, respectively, in a homogeneous out-of-plane applied magnetic field, $\mathbf{H}$. The SN contains an example of a single vortex, with a normal core of radius $\sim\xi$, and distribution of magnetic field around it characterized by $\Lambda$. When a current density $\mathbf{J}$ is sourced, the vortex will experience the Lorentz force, $\mathbf{F}_L$.}
	\label{fig:nanostripe}
\end{figure}

The numerical simulations performed in this work are all conducted on SN, such as exemplified in figure \ref{fig:nanostripe}. The SN have dimensions with lengths $L$, widths $w$, and thickness $d\ll\xi,\lambda$, where $\xi$ is the coherence length and $\lambda$ the magnetic field penetration depth of the superconducting state. For sufficiently large $H$, vortices form in the sample with a normal core of radius $\xi$ and penetration of the magnetic field up to a characteristic length of $\lambda$. In samples of our interest, being very thin, the effective penetration depth $\Lambda=\lambda^2/d$ by far exceeds the dimensions of the SN, such that the magnetic response of the superconductor is negligibly small compared to the applied magnetic field. Simulations of such SN are performed using the stationary (SGL) and time-dependent Ginzburg-Landau (TDGL) formalism. In the SGL approach we self-consistently solve the coupled equations
\begin{equation}
\left ( -i\nabla - \mathbf{A} \right )^{2}\Psi = \Psi \left ( 1-\left | \Psi \right |^{2} \right ),
\label{GL1}
\end{equation}
\begin{equation}
\vec{j}=-\kappa^{2} \nabla^2 \mathbf{A} = \frac{1}{2i}\left ( \Psi^{*}\nabla \Psi - \Psi \nabla \Psi^{*} \right ) - \left | \Psi \right |^{2}\mathbf{A}
\label{GL2}
\end{equation}
where $\Psi$ is the superconducting order parameter, $\mathbf{A}$ is the vector potential, and $\kappa =\Lambda/\xi$ is the effective Ginzburg-Landau parameter. We work with dimensionless units, where length is given in units of the temperature-dependent coherence length $\xi(T)=\xi$, the vector potential $\mathbf{A}$ in units of $c \hbar/2e\xi$, magnetic field $\vec{H}$ in units of the bulk upper critical field $H_{c2} = c\hbar / 2e\xi^{2}$, current in units of the GL current $j_{GL}=c\Phi_0/(8\pi^2\lambda^2\xi)$, and the order parameter $\Psi$ is normalized to its value in absence of applied field or sourced current ($\Psi_{0}$). We impose the Neumann boundary condition at the superconductor-insulator boundary at the lateral edges of the SN
\begin{equation}
\vec{n}\cdot \left(-i\nabla - \mathbf{A}\right)\Psi |_{boundary} = 0.
\label{boundary_nuemann}
\end{equation}
Along the length of the SN ($x$-axis) we enforce periodic boundary conditions for $\mathbf{A}$ and $\Psi$ (for the unit cell length $32\xi$, sufficient to capture the physics of interest in this work), of the form \cite{doria1989virial}
\begin{equation}
\mathbf{A}(x_{0} + L_{x}) = \mathbf{A}(x) + \nabla\chi_{f}(x)
\label{boundary_potential}
\end{equation}
\begin{equation}
\Psi(x_{0} + L_{x}) = \Psi(x)\exp\left[i \frac{2e}{\hbar c} \chi_{f}(x)\right],
\label{boundary_psi}
\end{equation}
where $\nabla\chi_{f}$ respects the gauge used for the magnetic field.
Equations \eqref{GL1} and \eqref{GL2} are solved numerically on a discretized Cartesian grid according to Ref. \cite{milovsevic2010ginzburg}, using the finite-difference method and the link-variable approach \cite{kato1993effects}, iteratively until convergence within a prespecified error is achieved. Then the supercurrent is calculated from the value of the order parameter and the vector potential (nearly entirely provided by the external magnetic field). With this method we obtain the vortex-row configuration-transition diagram as a function of $H$ and $w$ of the SN. 

The generalised time-dependent Ginzburg-Landau formalism \cite{ivlev1984electric,kramer1978theory} should instead be employed to properly study the dynamical properties of the superconducting condensate (with order parameter $\Psi(\textbf{r},t)$) in the presence of an external magnetic field $\textbf{H}$ (with vector potential $\textbf{A}$) and sourced current density $\textbf{J}$, given by
\begin{equation}
\begin{aligned}
	\tau _{GL}N(0)\frac{u}{\sqrt{1-(\Gamma|\Psi|)^{2}}}\left [ \frac{\delta\Psi}{\delta t} + i\tfrac{e^{*}}{\hbar}\varphi\Psi + \left ( \frac{\Gamma}{\sqrt{2}} \right )^{2}\frac{\delta |\Psi|^{2}}{\delta t}\Psi\right ]\\ 
	= -\left ( a + b|\Psi|^{2} \right )\Psi + \frac{\hbar^{2}}{2m^{*}}\left ( \nabla-ie^{*}\mathbf{A} \right )^{*}\Psi,
\end{aligned}
\label{tdgl1}
\end{equation}

\begin{equation}
	\nabla^{2} \varphi = \nabla\left [ Im \left \{ \Psi^{*}\left ( \nabla -i\textbf{A} \right ) \Psi\right \} \right ],
\label{tdgl2}
\end{equation}
where $a = \frac{\alpha}{2m^{*}\gamma}$, $b = \frac{\beta}{4m^{*2}\gamma^{2}}$, and $\Gamma = \frac{2\tau_{i}}{\hbar\sqrt{2m^{*}\gamma}}$. The Ginzburg-Landau order parameter relaxation time is $\tau_{GL}$; $N(0)$ is the density of states at the Fermi level; the parameter $u=5.79$ in conventional superconductors; $\varphi$ is the electrostatic potential; $\tau_{i}$ is the electron-phonon inelastic scattering time and $\alpha,\beta,\gamma$ are material parameters.
Equation \eqref{tdgl1} is solved coupled with the equation for the electrostatic potential (eq.\eqref{tdgl2}), using Neumann boundary conditions at all sample edges, except for the leads where sourced current is injected, where $\Psi=0$ and $\nabla\varphi=\pm J$. 
This theory is derived for dirty gapless superconductors, where Cooper-pair breaking occurs due to strong inelastic electron-phonon scattering, and the physical quantities $\Psi$ and $A$ must relax over a time-scale much longer than $\tau_{i}$. The distance over which an electric field can penetrate into the superconductor, and the length over which relaxation processes occur is given by the characteristic inelastic diffusion length $L_{i} = \sqrt{D\tau_{i}}$, where $D$ is the diffusion parameter proportional to the electronic mean-free path. In cases where $L_{i} <<\xi$, our simulations require very fine grid spacing (reflecting in consequently smaller time step in the used implicit Crank–Nicolson method) to yield physically correct results. 
In general, superconducting materials at $T$ close to the superconducting-to-normal transition temperature, $T_{c}$, satisfy the conditions for slow temporal and spatial variations ideally required for the applicability of the GL formalism. In the TDGL formalism, distances are given in units of $\xi(T)=\xi$; time in units of $\tau_{GL} = \frac{\pi\hbar}{8 k_{B}T_{c}(1-T/T_c)u}$; temperature is in units of $T_{c}$; the order parameter $\Psi$ in units of $\Delta(0) = 4k_{B}T_{c}u^{1/2}(1-T/T_c)^{1/2}/\pi$; $\varphi$ in units of $\varphi_{GL} = \hbar/e^{*}\tau_{GL} $; vector potential A is scaled to $A_0 = H_{c2}\xi$ and current density to $J_{0} = \sigma_n\varphi_0/\xi$. 
The simulations are performed irrespective of the temperature $T/T_c$, all physical quantities are scaled and normalised by reference quantities at a given temperature.
Finally, in our analysis we do not consider the heating effects, which would require coupling of the TDGL formalism to the thermal balance equation \cite{jelic2016velocimetry}. This is justified for samples that are very efficiently thermally coupled to the substrate and/or thermal bath (i.e. have large heat transfer coefficient).

\section{Results}
In what follows, using SGL and TDGL simulations, we study how confinement forces in narrow SN affect the stationary vortex configurations and their dynamics, under $H$ and a sourced dc current density, $J$. 

\subsection{Equilibrium vortex configurations}
\begin{figure}[t]
	\centering
	{\includegraphics[width=\linewidth]
		{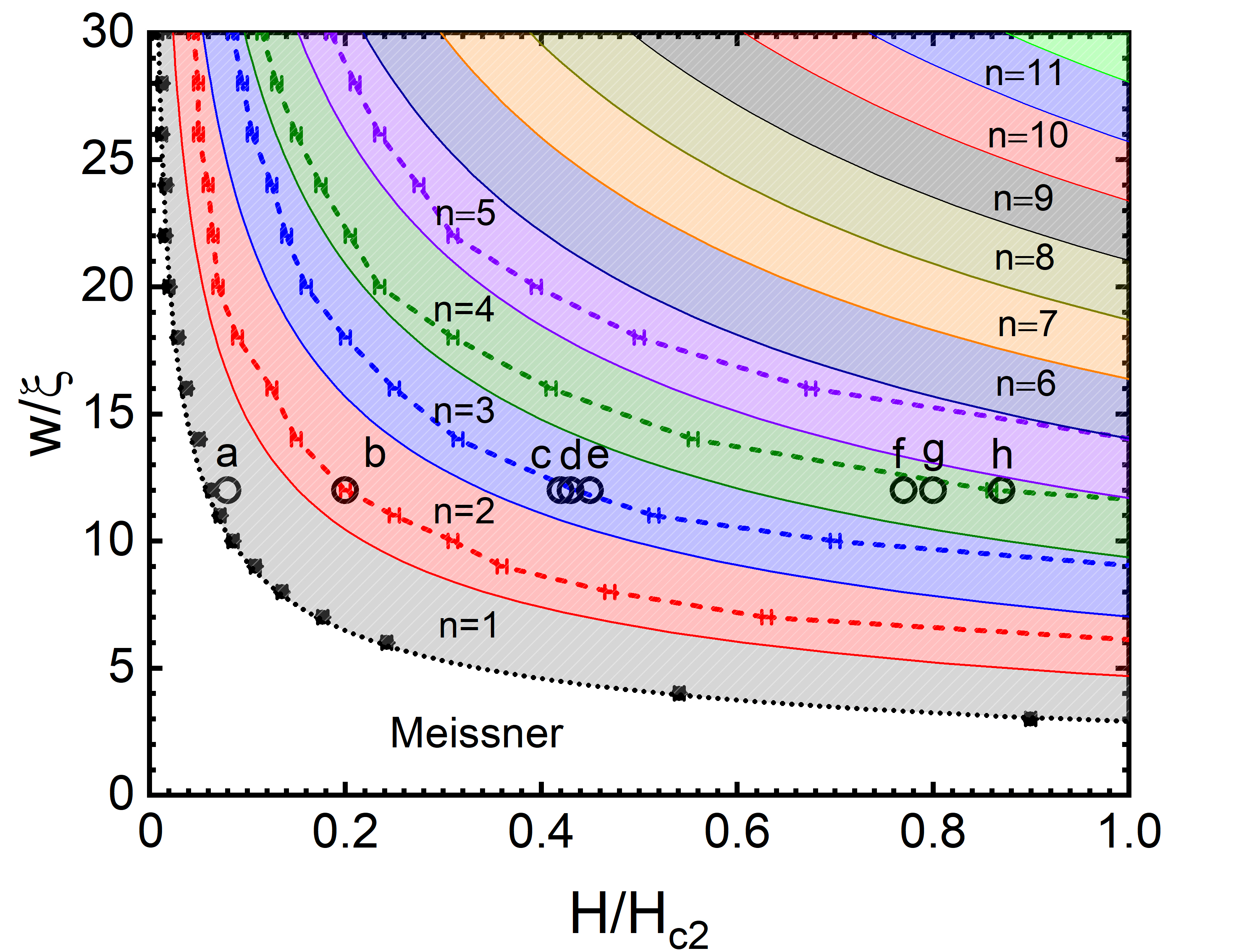}}
	\caption{Equilibrium vortex-row phase diagram, plotting the SN width (in units of $\xi$) as a function of the applied magnetic field intensity (in units of $H_{c2})$, for different numbers of formed vortex rows($n$). Simulations have been done using the SGL approach with periodic boundary conditions along the length, with unit cell length of $L=32\xi$. The dashed lines denote the threshold for the formation of an additional vortex row (here shown up to $n=5$). The coloured regions represent the approximated regions for $n>1$, delimited by solid lines given by expression $H_{row}/H_{c2}$ = $\frac{\pi n^2 \xi^2}{\sqrt{3}w^2}$. Circles, labeled $a-h$, relate to the vortex configurations shown in figure \ref{fig:row_examples}. Black dotted line corresponds to the analytical expression $H_0/H_{c2} = K\frac{\pi^{2}\xi^2}{2w^2}$ \cite{clem1998paper}, with $K=1.7$ \cite{stan2004critical}.}
	\label{fig:phase_diagram}
\end{figure}

\begin{figure}[t]
	\centering
	{\includegraphics[width=\linewidth]
		{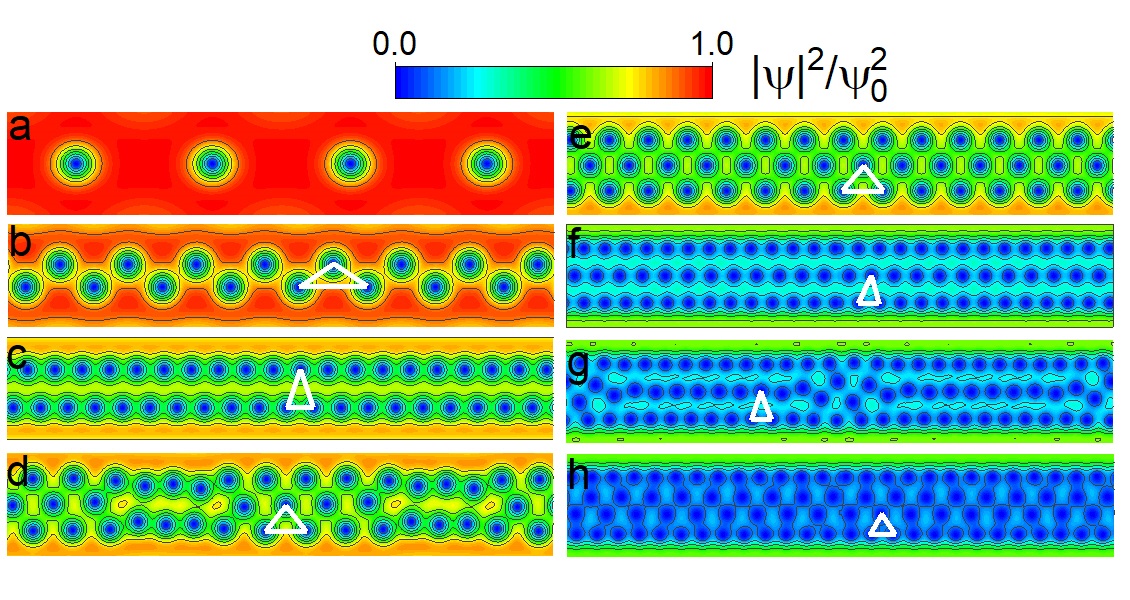}}
	\caption{Calculated vortex configurations plotted as Cooper-pair density for the ground state of a SN of width $w = 12\xi$,  in a periodic cell of $L=64\xi$, at different applied $H/H_{c2}$ values: (a) 0.08; (b) 0.20; (c) 0.42; (d) 0.43; (e) 0.45; (f) 0.77; (g) 0.80; (h) 0.87 (cf. figure \ref{fig:phase_diagram}). Panels (b), (d) and (g) depict the vortex states at the nucleation of a second, third and fourth row, respectively. Panels (c), (e) and (h) show the most lattice-like packing conditions for two, three and four vortex rows, respectively. White lines, connecting the cores of three neighbouring vortices, illustrate the deformation of the Abrikosov lattice \cite{abrikosov2004nobel} in the SN. Color bar denotes the values of the Cooper-pair density shown in the panels. Each depicted configuration is indicated in figure \ref{fig:phase_diagram} with an open dot and is labelled accordingly.}
	\label{fig:row_examples}
\end{figure}

We start by producing the vortex row phase diagram using the SGL approach, showing the conditions for the formation of a number of vortex rows, $n$, as a function of $H$ and $w$ of the SN. Each dashed curve in the diagram shown in figure \ref{fig:phase_diagram}, plotting the width of the SN versus $H$, represents the appearance of the $n^{th}$ vortex row ($n=1-5$) in the ground state of the system as magnetic field is increased. Examples of corresponding vortex configurations for a SN of $w = 12\xi$ for different $H$ intensity are shown in figure \ref{fig:row_examples}, corresponding to the pinpointed dots (labelled a-h) in figure \ref{fig:phase_diagram}.
To identify the threshold $H$ for the transition to the vortex row configuration with a higher $n$, the ground states were first obtained for each SN at different $H$; then the spatial distribution of the superconducting order parameter $|\Psi|^2$ has been plotted (similar to figure \ref{fig:row_examples}) and carefully analysed, focusing on the geometrical interpretation of the vortex configuration. In the SGL approach adopted in our simulations, the SN was considered periodic along its length, with a unit cell of $L = 32\xi$. Several checks, carried out by extending the unit cell length till $80\xi$, have confirmed all following results.

In a SN, the early theoretical works \cite{maksimova1998mixed, clem1998paper} have shown that the magnetic field at which the surface barrier is suppressed and a single vortex can be stable in the SN is $H_0/H_{c2}=\pi^2\xi^2/2 w^2$. The subsequent experimental observations of vortex penetration fields by Stan \textit{et al.} \cite{stan2004critical} have shown a very good agreement with latter expression, up to a multiplying constant $K$. Our numerical data (black dots in figure \ref{fig:phase_diagram}) reconfirm that finding, as vortex penetration fields were found to nearly ideally match the same functional dependence on $w$, with a multiplying constant $K=1.7$.

The approximate criteria for further reconfiguration of the vortex states and appearance of additional vortex rows can be obtained in the following way. We consider the Abrikosov triangular lattice, with the lattice parameter $a = 1.075\sqrt{\phi_{0}/H}$. The vortices are arranged in a body-centered hexagonal lattice, and so the Wigner-Seitz unit cell is hexagonal with a unit area per flux quantum of $A = \frac{\sqrt{3}}{2}a^{2}$. For a narrow SN, to accommodate $n$ rows of vortices, the spacing, $w_v$, between vortex rows must obey the inequality $w_v\leq w/n$. Using the previous expression for the Abrikosov vortex density, we substitute $A=\frac{\sqrt{3}}{2}w_v^2=\frac{\sqrt{3}w^2}{2n^2}$ to obtain the zeroth order approximation for the threshold magnetic field required for the formation of new rows, yielding $H_{row}/H_{c2}$ = $\frac{\pi n^2\xi^2}{\sqrt{3}w^2}$. Those approximate threshold $H$ values are shown in figure \ref{fig:phase_diagram} by the solid lines delimiting different coloured regions, indicating transitions between states with different number of vortex rows. In general, the behavior of threshold $H$ found using SGL simulations agrees well with the formula prediction. The values were however mostly higher than the approximate ones, which is attributed to the role played by the edge barriers for vortex entry and exit (varying, depending on $w$ and $H$). In addition, the rearrangement of the vortex lattice with every vortex penetration is not taken into account in latter basic analytical formula. Note that such effects of the vortex-vortex interactions and interactions with the edge Meissner currents (causing the confinement force), dominate the formation of the vortex configurations in narrow SNs and present the main point of interest in this work.

\begin{figure}[b]
	\centering
	{\includegraphics[width=\linewidth]
		{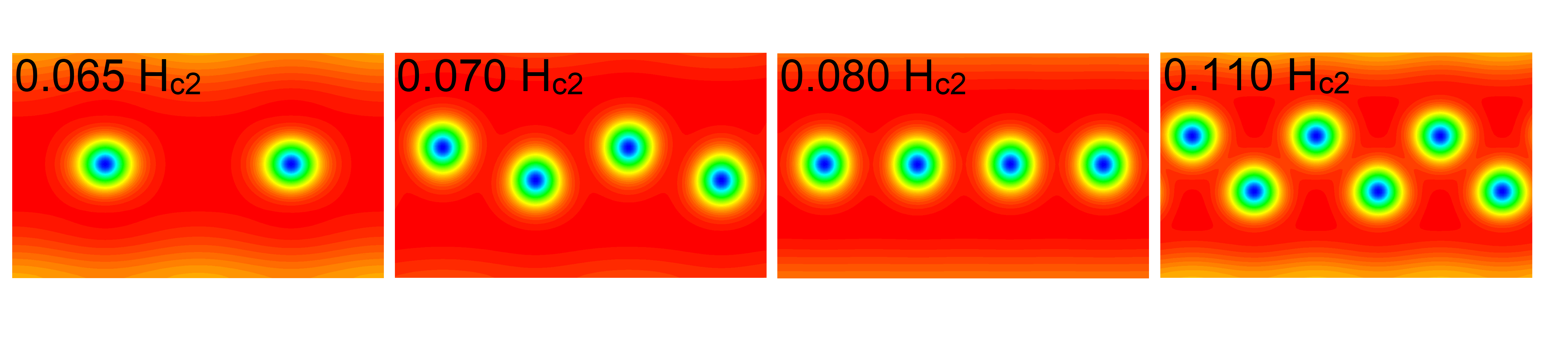}}
	\caption{Exemplified re-entrant transition between the one- and two-row vortex state in a SN of width $w=20\xi$, caused by the competition between the confinement imposed by Meissner currents and the vortex density, while both changing with increasing $H$ (value indicated inside the panels).}
	\label{fig:reconfiguration}
\end{figure}

For a SN of $w=12\xi$, we show different vortex-row configurations in figure \ref{fig:row_examples}, as formed in the ground state at different $H$ values (marked by open dots in figure \ref{fig:phase_diagram}). After formation and growing of the first vortex row population (figure \ref{fig:row_examples}-a), increasing $H$ vortices rearrangement into a closely packed ``zig-zag" state (figure \ref{fig:row_examples}-b). This close packing is emphasised by a white triangle progressively deviating from the equilateral shape expected in the Abrikosov vortex lattice, with rise of $H$. Obviously, in this state the Meissner currents will exert a strong repulsive and confining force on the vortices from the SN edges (i.e. strong Bean-Livingston edge barrier \cite{bean1964surface}), resulting in a vortex spacing far smaller than the above rough analytical estimates (leading to the solid lines in figure \ref{fig:phase_diagram}).

Starting from the one row configuration (figure \ref{fig:row_examples}-a), raising $H$ and $n$ further, strengthens the relevance of the vortex-vortex interaction forces on the resulting vortex configuration, which will increase separation between the two rows (figure \ref{fig:row_examples}-c). At this point, we observe that additional vortices in the SN cannot uniformly balance the aforementioned competing force in the entire SN, leading to a local rearrangement of the vortex lattice to three rows (figure \ref{fig:row_examples}-d). Only with further increasing field and having enough vortices in the SN the full three-row state is formed (figure \ref{fig:row_examples}-e; notice a nearly ideal triangular lattice formed). For the considered width of the SN, the state with 3 vortex rows persists to a much larger field due to quantum confinement, such that vortices very strongly overlap in a closely packed structure (figure \ref{fig:row_examples}-f). Nevertheless, in the vicinity of the bulk upper critical field a fourth row forms, first locally (figure \ref{fig:row_examples}-g) and eventually in the entire SN (figure \ref{fig:row_examples}-h), before superconductivity is destroyed. No further rows of vortices can form at higher field and the existing vortex rows increasingly overlap until the normal state is established. 

\begin{figure}[t]
	\centering
	{\includegraphics[width=0.9\linewidth]
		{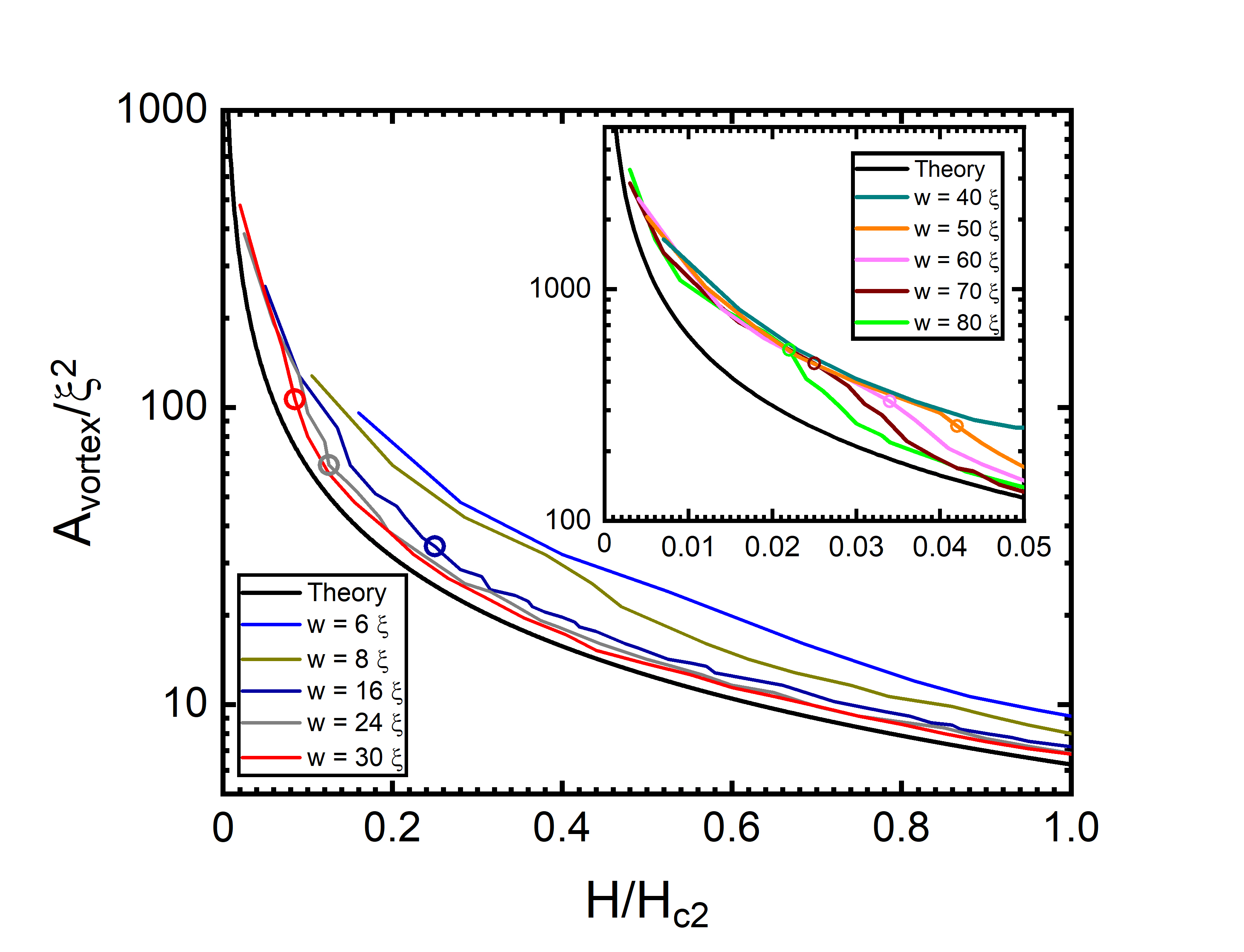}}
	\caption{Area of the Wigner-Seitz unit cell containing a single vortex, as a function of the applied magnetic field, for SN of width $w=6-30\xi$, and $w= 40-80\xi$ in the inset for small $H$ values. The analytical expression for the Abrikosov vortex lattice (AVL) area, $\frac{A}{\xi^2}$ = $2\pi\frac{H_{c2}}{H}$, is plotted as a black line. Open dots in each curve indicate the $H$ intensity for formation of the third vortex row, above which the curves progressively approach the AVL expression, upon increasing the $w$.}
	\label{fig:vortex_density}
\end{figure}
 
We reiterate that the transitions between rows of vortices and the final arrangement of vortices in the lattice are strongly affected by competition of the two forces, both dependent on $H$. As the magnetic field is increased the edge Meissner current also increases, up to the penetration of new vortices, while every new vortex changes the landscape of the vortex-vortex interactions in the SN. As exemplified in figure \ref{fig:reconfiguration} for $w=20\xi$, this nontrivial balance of competing forces can lead to a re-entrant behavior in terms of the number of vortex rows formed. 
In such cases, the zig-zag instability of the vortex row can be "cured" back into a single row by the increasing Meissner currents, as the lateral confinement forces grow with increasing magnetic field. As $H$ is increased further the additional penetrating vortices tip the scale in favor of vortex interactions and a definite reconfiguration into a state with two rows form. 
This re-entrant behavior has been observed for nearly all considered SN widths in the range $w=20-60\xi$. In such cases we have taken the first onset of the zig-zag instability to mark the $n\rightarrow n+1$ transition in figure \ref{fig:phase_diagram}. Moreover, this range of widths, where such strong edge effects are detected, marks the crossover from the quasi-1D to a 2D film-like behavior.

As the magnetic field is increased, vortices penetrate the SN of different width and vortex rows are formed; and a gradual evolution from a quasi-1D row pattern into a 2D vortex lattice is expected. To evaluate such crossover, we calculated the average area occupied by a single vortex as a function of $H$ in all the states found, and compared to the expected behavior of the Abrikosov vortex lattice area. The strong confinement in narrowest SN \cite{kimmel2019edge,bean1964surface} dominates the vortex-vortex interaction, leading to compression of vortices into fewer vortex rows and consequently larger average area per vortex. This can be seen in figure \ref{fig:vortex_density} for $w\leq 8\xi$. As the width of the SN is made larger, the confining force from the edge current (at a given $H$) becomes less dominant with respect to the vortex-vortex interaction, resulting into progressively closer agreement with the expected behaviour of a triangular vortex lattice \cite{abrikosov2004nobel}. This tendency is clearly visible upon formation of the third vortex row (cf. figure \ref{fig:vortex_density}). 

\begin{figure}[t]
	\centering
	{\includegraphics[width=0.8\linewidth]
		{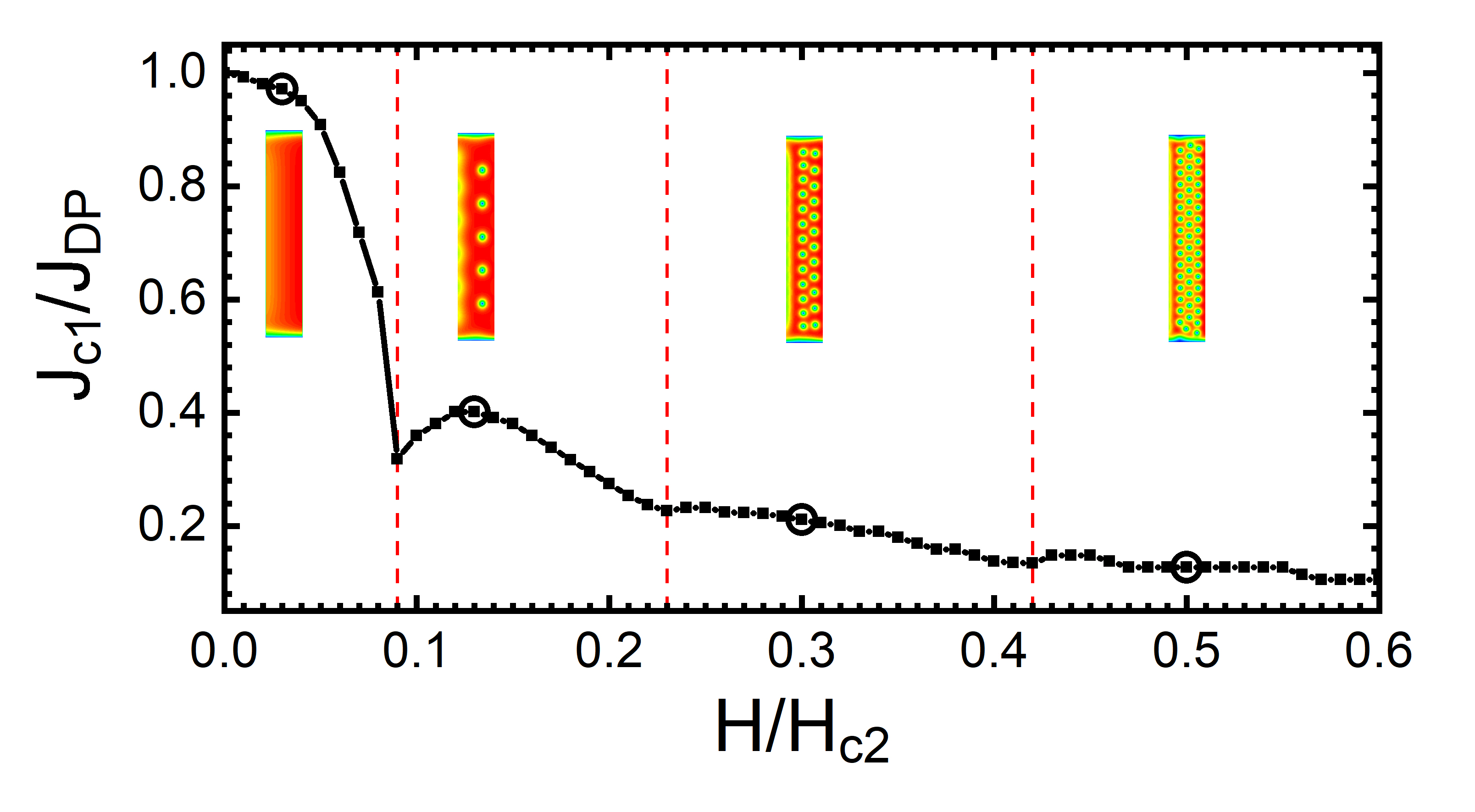}}
	\caption{First critical current density normalised to $J_{DP} = 0.385 J_{GL}$ as a function of the applied magnetic field normalised to $H_{c2}$, for a SN of width $w = 12\xi$, obtained using the TDGL approach. Vertical red lines mark the transition to 1, 2 and 3 vortex-row states, at a magnetic field of $H/H_{c2}$ = 0.09, 0.23, 0.42, respectively. Insets illustrate the vortex row configurations at selected magnetic fields (marked by open dots) for applied current density just under the critical one.}
	\label{fig:Jc_vs_H}
\end{figure}

\subsection{Vortex dynamics under sourced current}
All above results have been obtained in the stationary case, where no current is sourced to the SN. A sourced current may change the stationary states or induce vortex dynamics specific to the nanoconfined regime. In what follows, we examine those non-equilibrium effects, through TDGL simulations of time-dependent processes.\\
When a small transport current is applied along a SN under an applied magnetic field $H\geq H_{c1}$, the present vortices experience a push across the SN due to the Lorentz-type of force ($\propto\mathbf{J}\times \mathbf{H}$). As the applied current density is increased, vortices will continue to shift across the SN, finally leaving the stripe for sufficiently large Lorentz force. This defines the first critical current density ($J_{c1}$), for which vortices are able to overcome the edge barrier \cite{kimmel2019edge,bean1964surface} and start to cross the SN continuously, nucleating on one side, moving across the SN, and exiting at the opposing edge. The critical current depends on magnetic field $H$ for a given width $w$ of the SN. The first critical current density as a function of $H$, $J_{c1}(H)$, for a SN of $w = 12\xi$, is shown in figure \ref{fig:Jc_vs_H}. A commensurate effect is observed between $H$ and $n$, where minima in the curve correspond to the transition to a state with an additional vortex row. Previous works have reported similar behaviour using different theoretical approaches \cite{mawatari1994critical,carneiro1998equilibrium}, including a comprehensive study using the TDGL approach \cite{vodolazov2013vortex} and revealing the relation between $n$, $w$ and applied magnetic field \cite{kimmel2019edge}. The increase in $J_{c1}$ from local minima as the applied $H$ field is further increased, is caused by the competition between vortex-vortex interactions and the confinement from the SN edge. After a local minimum, when a new row is formed, the vortex-vortex interactions are strong and the confining edge currents producing an entry/exit barrier are reduced and more easily overcome with lower applied currents. As $H$ is increased, the induced Meissner currents at the edge increase \cite{vodolazov2003vortex}, reinforcing the edge barrier. The vortex-row phase diagram in figure \ref{fig:phase_diagram} can be used to predict the transition field value, where local minima occur in $J_{c1}(H)$ curves - which is an experimentally verifiable feature. Note however that the threshold fields for formation of new rows in presence of applied current are somewhat different from the ones presented in figure \ref{fig:phase_diagram}, since the Lorentz push exerted by the current effectively increases the confinement experienced by vortices prior to the onset of their motion.

\begin{figure}[t]
	\centering
	\includegraphics[width=\linewidth]{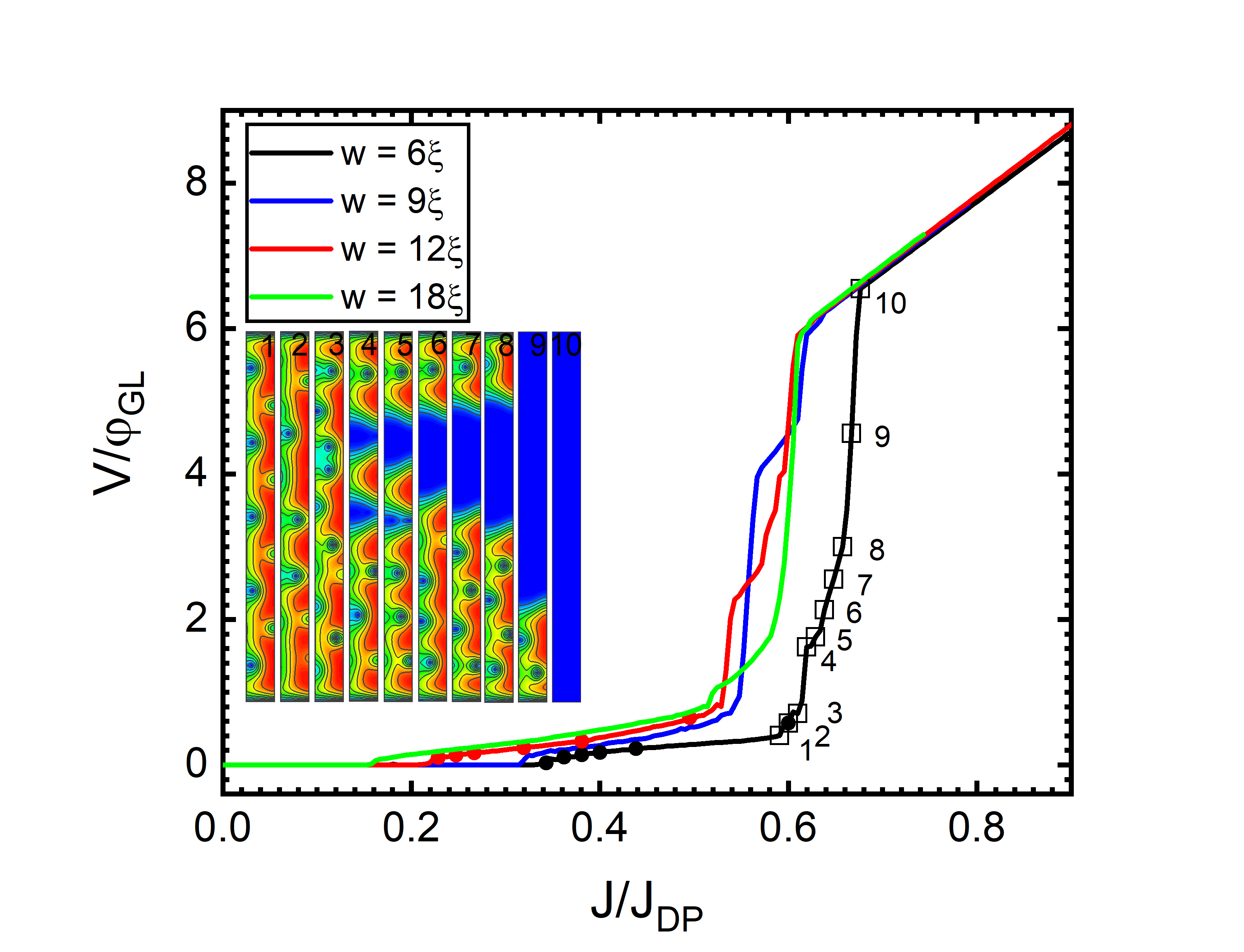}
	\caption{Normalised voltage drop as a function of the normalised current density, for SNs of width $w$ = 6, 9, 12, and 18$\xi$, at magnetic field $H = 0.25 H_{c2}$. The black and red dots (for $w$ = 6 and 12$\xi$ respectively) mark the values of current density at which analysis of the modulation frequency spectra is presented in figures \ref{fig:FFT6} and \ref{fig:FFT12}. Inset: Snapshots of the Cooper-pair density for SN of $w$ = 12$\xi$, numbered 1-10, indicated by open red squares.}
	\label{fig:IVs}
\end{figure}
The TDGL approach has allowed us to also simulate the voltage-current density (V-J) characteristics of SNs, presented in figure \ref{fig:IVs} for stripes with $w$ = 6, 9, 12, 18$\xi$, under an applied magnetic field $H = 0.25H_{c2}$. Analysis of the V-J characteristics reveals a number of features related to different resistive regimes in each curve. At low $J$ values, stationary vortices are shifted to a new position across the SN due to the Lorentz force produced by the sourced current, so the resulting voltage drop and resistance remain zero. An example of such can be seen in figure \ref{fig:IVs} (for $w = 12 \xi$) from the states labelled 1 and 2. 
When $J_{c1}$ is reached, vortices cross the SN and their perpetuous motion leads to a finite resistivity value. Snapshots of this flux-flow regime can be seen from the states labelled 3 and 4 in figure \ref{fig:IVs}.
With further increasing $J$ and in presence of vortex-vortex interaction forces, a SN in the dissipative state exhibits flux-flow instability, where vortex cores interact during dynamics and ordered lattice structure is lost during motion (state labelled 5 in figure \ref{fig:IVs}). At even higher $J$, vortices align during motion, in a slip-streamed geometry (vortices tailgate, i.e. subsequent vortices, crossing the SN, move in the wake of the previous vortex \cite{vodolazov2007rearrangement,embon2017imaging}), before a Langer-Ambegaokar phase slip \cite{langer1967intrinsic} occurs across the SN. The normal area covered by the phase slip grows laterally with further increasing $J$, and additional steps in the V-J curve appear with every slip-stream being merged with the growing phase-slip, as seen in the states labelled 6-10 in figure \ref{fig:IVs}.
When $J$ reaches roughly 0.65$J_{DP}$, the SN transitions to a fully normal state, with linear ohmic behavior. Similar V-I curves have been observed both numerically \cite{cadorim2020ultra} and experimentally for Nb-C microstrips, fabricated using focused-ion-beam-induced deposition \cite{dobrovolskiy2020ultra}.
\begin{figure}[t]
	\centering
	{\includegraphics[width=\linewidth]{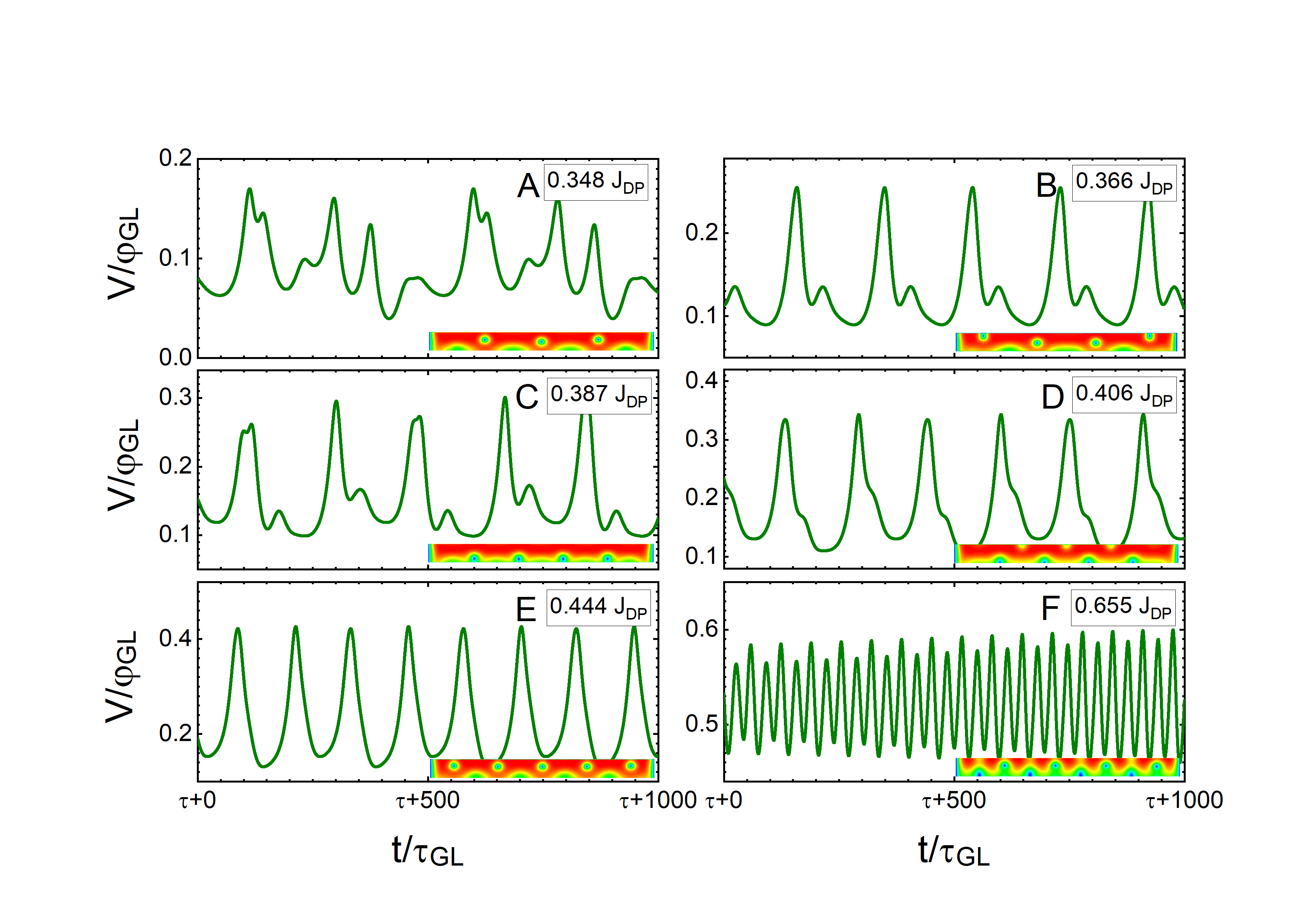}}
	\caption{Normalised voltage drop as a function of time (normalised to $\tau_{GL}$), for a SN of $w = 6\xi$ under a magnetic field of $H = 0.25 H_{c2}$ sourced with different current densities of: (A) 0.348$J_{DP}$, (B) 0.366$J_{DP}$, (C) 0.387$J_{DP}$, (D) 0.406$J_{DP}$, (E) 0.444$J_{DP}$, and (F) 0.655$J_{DP}$. Each panel contains an illustrative snapshot of the spatial distribution of the Cooper-pair density during the dynamics.}
	\label{fig:modulations6xi}
\end{figure}
\begin{figure}[t]
	\centering
    	\includegraphics[width=\linewidth]{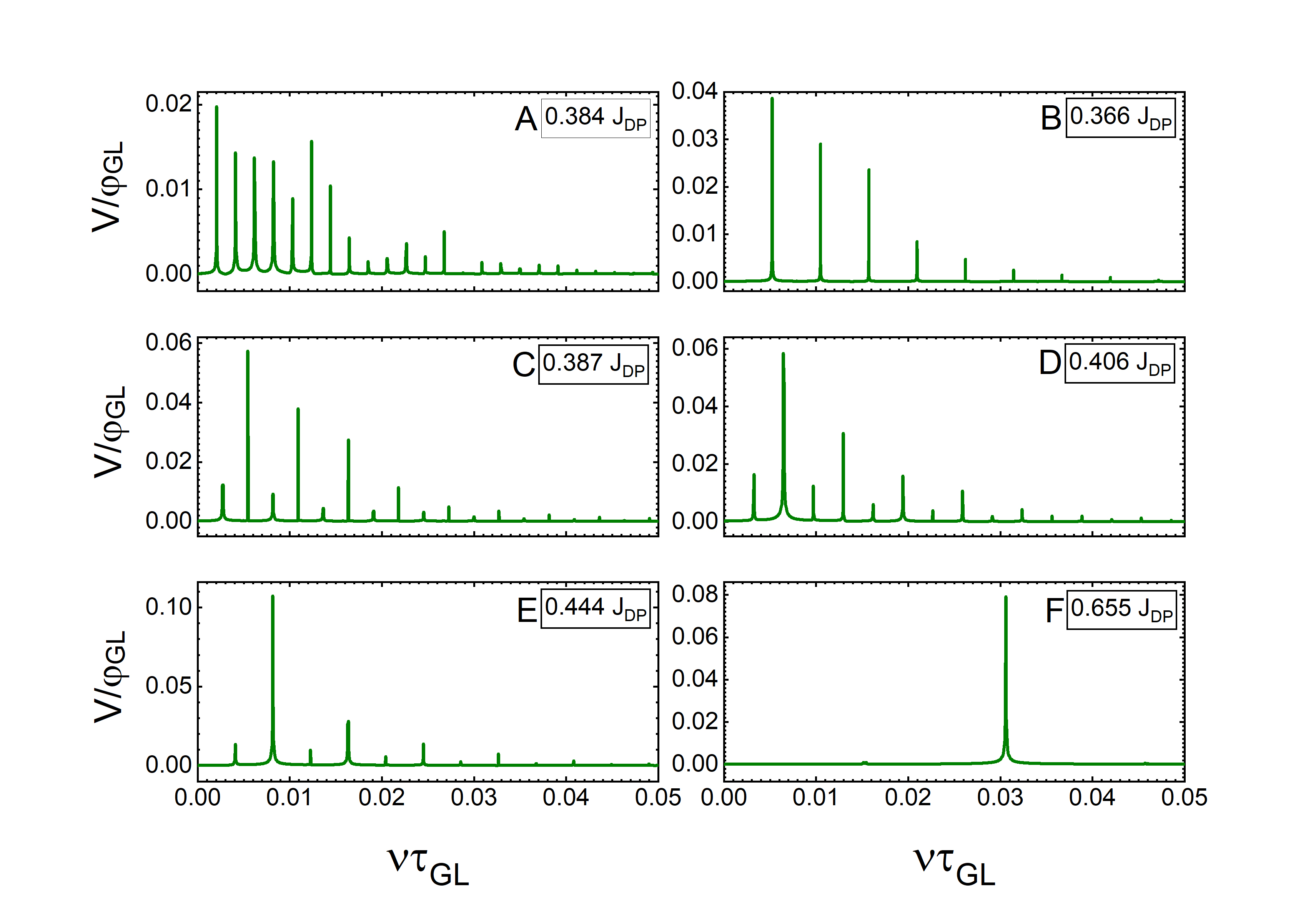} 
	\caption{Spectra of modulation frequencies $\nu$ (normalised to $\tau_{GL}^{-1}$) of the temporal voltage signals shown in figure \ref{fig:modulations6xi}.}
	\label{fig:FFT6}
\end{figure}

\begin{figure}[ht]
	\centering
	{\includegraphics[width=\linewidth]{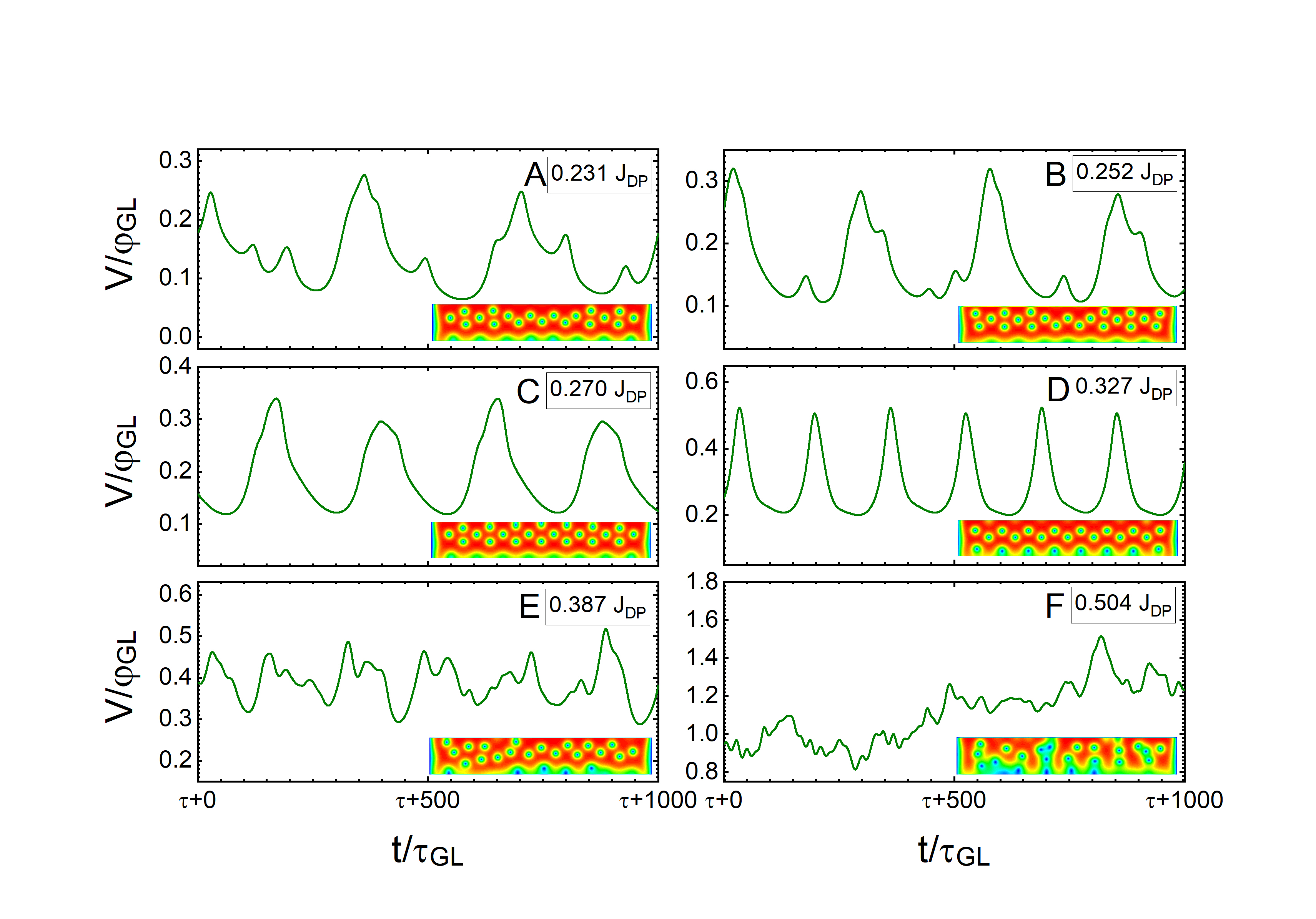}}
	\caption{Normalised voltage drop as a function of time (normalised to $\tau_{GL}$), for a SN of $w = 6\xi$ under a magnetic field of $H = 0.25 H_{c2}$ sourced with different current densities: (A) 0.231$J_{DP}$, (B) 0.252$J_{DP}$, (C) 0.270$J_{DP}$, (D) 0.327$J_{DP}$, (E) 0.387$J_{DP}$, and (F) 0.504$J_{DP}$. Each panel contains an illustrative snapshot of the spatial distribution of the Cooper-pair density during the dynamics.}
	\label{fig:modulations12xi}
\end{figure}

\begin{figure}[ht] 
	\centering
    	\includegraphics[width=\linewidth]{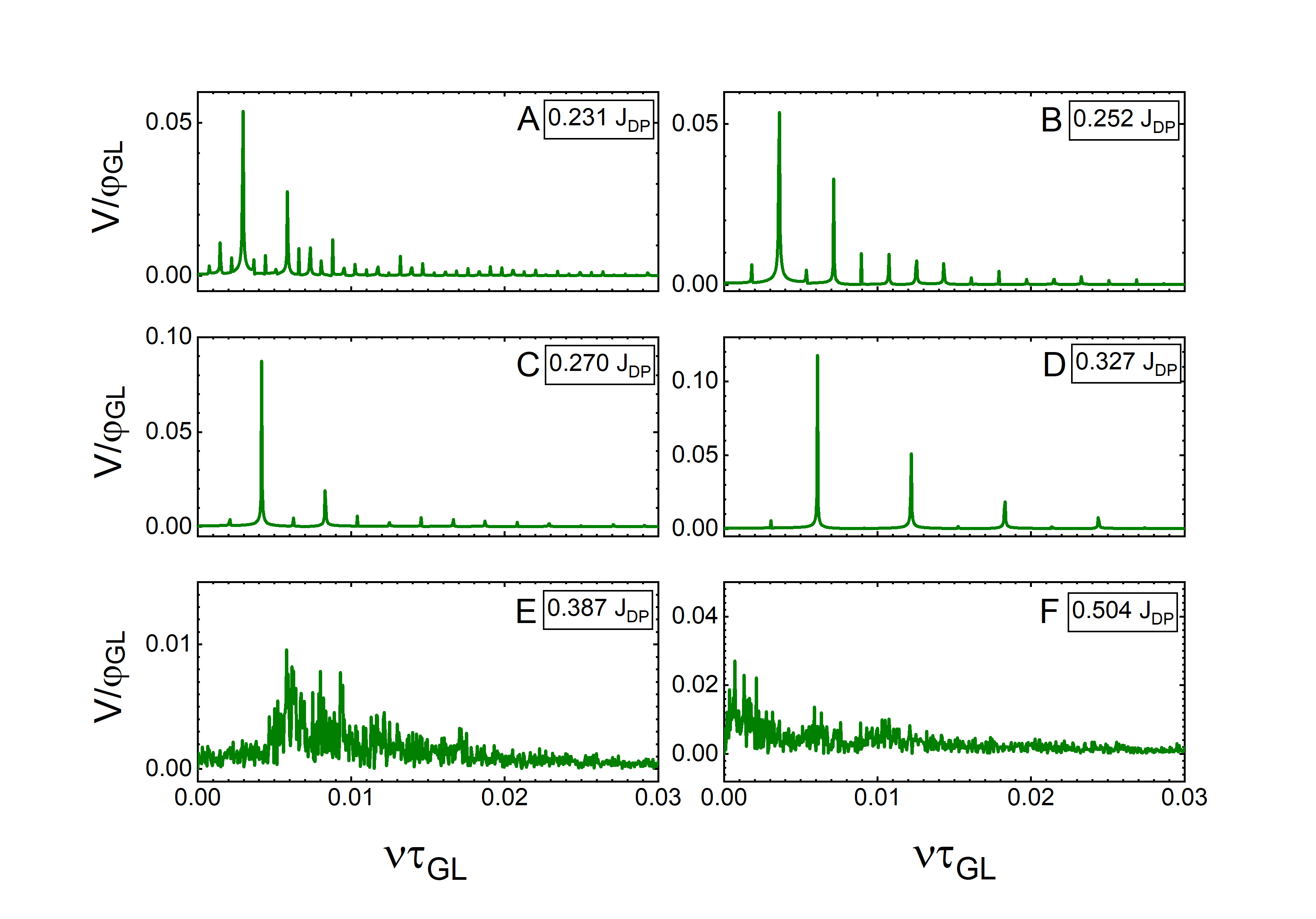} 
	\caption{Spectra of modulation frequencies $\nu$ (normalised to $\tau_{GL}^{-1}$) of the temporal voltage signals shown in figure \ref{fig:modulations12xi}.}
	\label{fig:FFT12}
\end{figure}

Next we discuss how the observed vortex crossings modulate the voltage drop across the SN, and how synchronous and asynchronous crossings affect the spectrum of frequencies as a consequence of those modulations. In figures \ref{fig:modulations6xi}-\ref{fig:FFT12} we show the voltage as a function of time, $V(t)$, for different applied current $J$, and their corresponding spectra of frequencies (obtained by Fourier transform of $V(t)$) for SNs of $w=6\xi$ and $w=12\xi$, under an applied field $H=0.25H_{c2}$. In each case, we have first used the TDGL approach to find the ground states for each SN at the given magnetic field, then we swept the $J$ from 0 up to $\simeq J_{DP}$, in sufficiently small steps (typically $\simeq 0.025 J_{DP}$). At each current step the simulation has been left to run for sufficiently long time so that a dynamic equilibrium has been reached (typically up to $t = 5\times 10^3$ $\tau_{GL}$), before recording data. The so obtained $V(t)$ and the spatial distribution of the superconducting order parameter at each time step, are used to produce figures \ref{fig:modulations6xi} and \ref{fig:modulations12xi}. 
In the dissipative state, $V(t)$ raises as vortices move across the SN, with maxima corresponding to the exit of a vortex and minima to an entry of a vortex \cite{jelic2015stroboscopic,jelic2016velocimetry}, leading to modulations of $V(t)$ for both SN considered (figures \ref{fig:modulations6xi} and \ref{fig:modulations12xi}). 
Considering the SN of $w=6 \xi$, sourced with the lowest current causing the vortex crossing ($J=0.348 J_{DP}$ in this case), $V(t)$ shows evidence of asynchronous vortex dynamics, with several distinct features having a periodicity of 486 $\tau_{GL}$.
Even though the vortices are not crossing in synchronised rows, there is a quasi-synchronised behaviour manifesting in the repetition of vortex crossings in a given dynamic configuration. As $J$ is increased from 0.384$J_{DP}$ to 0.406$J_{DP}$ (panels A-D in figure \ref{fig:modulations6xi}) the modulations in the voltage evolve, and the number of modulations caused by quasi-synchronous crossings reduces. Finally beyond $J = 0.444 J_{DP}$ (panel E) there is only one mode that repeats periodically, i.e. vortex dynamics becomes fully synchronous, and accelerates with further increasing the current (panel F). The relative spectra of frequencies for $V(t)$ are shown in figure \ref{fig:FFT6}, in panels labelled correspondingly to panels of figure \ref{fig:modulations6xi}. The repetitive modes of vortex crossings within the particular dynamic configuration lead to peaks at specific frequencies. As the current density is increased, the spectra show the evolution to a single peak, corresponding to the frequency of $0.03\tau_{GL}^{-1}$. 

A similar analysis for the SN with $w = 12\xi$ is shown in figure \ref{fig:modulations12xi}, for the reason that a wider nanostripe allows formation of multiple vortex rows in the ground state. Panel A of figure \ref{fig:modulations12xi} shows $V(t)$ at $J = 0.231 J_{DP}$, when vortices start dissipatively crossing the SN, in a quasi-synchronous fashion. As $J$ is increased to $0.327 J_{DP}$, the vortex crossings become increasingly synchronised (panels B-D). However, at $J = 0.387 J_{DP}$ the flux-flow instability sets in (panel E), and causes an increasingly chaotic behavior as $J$ is increased to $J = 0.504 J_{DP}$ (panel F). In this regime, the apparent chaotic behavior is caused by the competition between the standard vortex-vortex repulsion and the effective attractive core-core interaction due to preferential tailgating at large vortex velocities, interchanging their dominance on each vortex during the collective dynamics. For $J > 0.52J_{DP}$ a phase slip occurs that will grow as applied current density is raised (shown in figure \ref{fig:IVs}, states labelled 6-10), and the remaining vortices are crossing the stripe in tailgated rows. Such case of tailgated vortices causes periodic modulations $V(\tau)$, however, at such high $J$ this regime is unstable and therefore not considered in the proceeding discussion. So we only consider the region of strictly flow-flow during the discussion of synchronised vortex crossings.\\
The spectra of frequency modulations (figure \ref{fig:FFT12}) show analogous behaviour to that of the narrower SN discussed previously. At low $J$, when crossings are quasi-synchronous, we see many mode contributions (i.e. few dominant peaks accompanied with many additional smaller peaks). As $J$ is increased and synchronicity improves, the smaller contributions disappear, and the frequency component with the largest contribution is strengthened. However, at the onset of the flux-flow instability ($J = 0.387J_{DP}$) we observe a broad contribution centred around frequency $\nu=0.06\tau_{GL}^{-1}$ (corresponding to the median frequency of crossing of the vortex lattice as a whole, with many individual asynchronous crossings superimposed). At $J = 0.504 J_{DP}$ (panel F) the spectrum loses any order, corresponding to the chaotic behaviour of vortex crossings.
Vortices continuously crossing the SN, will cause oscillations in the electric and magnetic fields, leading to detectable emission of electro-magnetic radiation \cite{dobrovolskiy2018microwave,bulaevskii2006electromagnetic,dolgov2000transition}. The crossing of a single vortex releases a very small amount of energy, whereas multiple vortices moving coherently will emit a significant (and more easily detectable) amount of energy \cite{dolgov2000transition}. In a coherently moving lattice of vortices, periodic vortex crossing in the SN will cause emission of radiation at a frequency of $\omega = 2\pi v/a$ (washboard frequency) and at harmonics $\omega = 2\pi m v/a$ ($m=2,3...$), where $v$ is the vortex speed and $a$ is the lattice spacing (i.e., the distance between two parallel adjacent rows in our case) along the direction of motion \cite{bulaevskii2006electromagnetic}. The highest frequency emitted cannot exceed $\Delta/\hbar$, where $\Delta$ is the superconducting gap of the SN. The theoretically predicted existence of radiation has been experimentally confirmed \cite{dobrovolskiy2018microwave}. 
\begin{figure}[t] 
	\centering
    	\includegraphics[width=\linewidth]{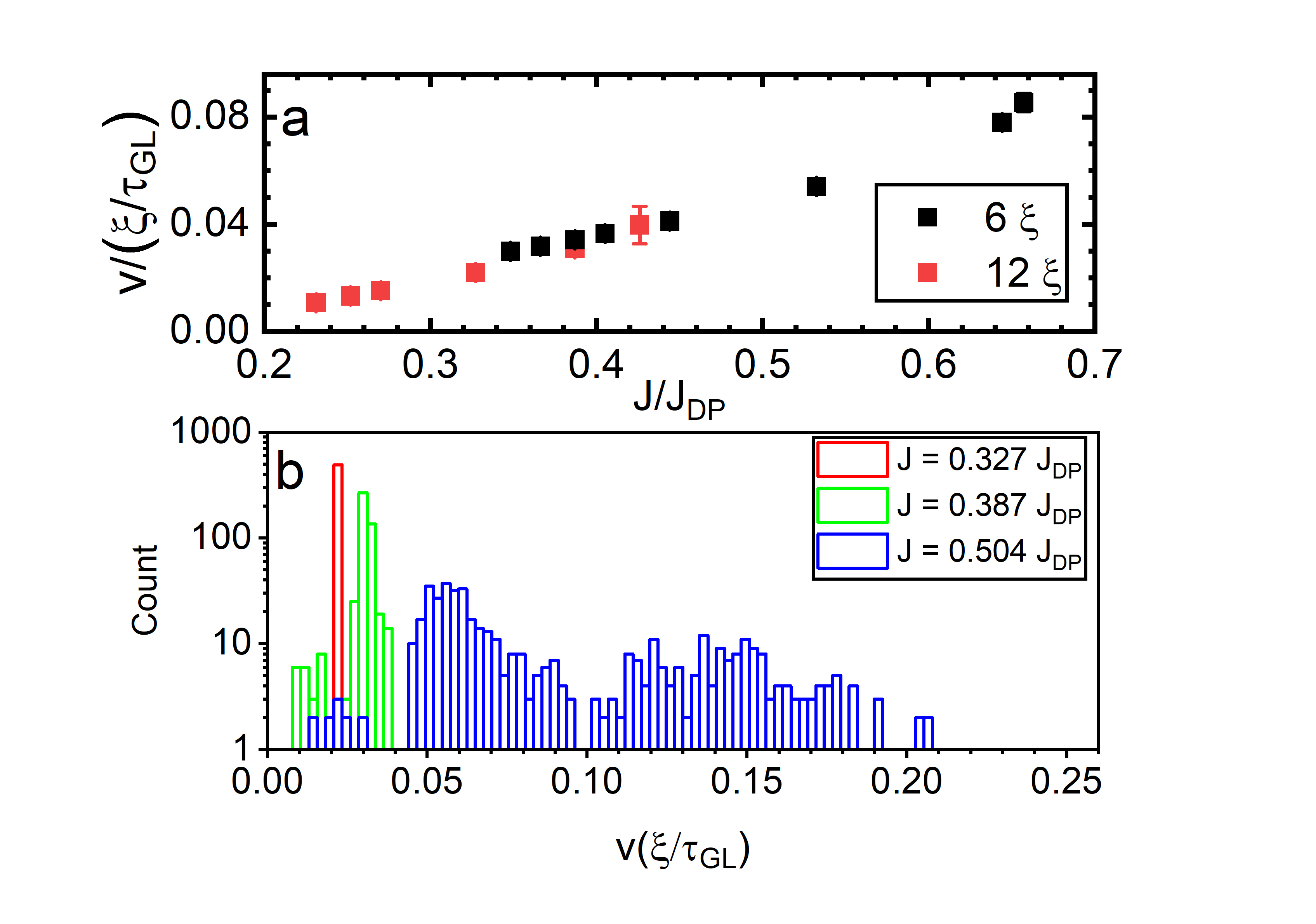} 
	\caption{Panel a - Normalised average vortex velocity in units of $\xi/\tau_{GL}$ versus the normalised current density applied to SN of width $6\xi$ and $12\xi$ under an applied magnetic field of $H = 0.25H_{c2}$. Panel b - Histogram of vortex velocities for different values of $J$ relating to different vortex crossing regimes.}
	\label{fig:vortex_velocity}
\end{figure}

Results of our simulations of the vortex velocity as a function of applied current density are shown in figure \ref{fig:vortex_velocity}a. They evidence a linear dependence at lower values of $J$, for both above-considered narrow SNs (similar to behaviour seen in \cite{cadorim2020ultra,berdiyorov2009kinematic}). However, when $J$ is increased to intermediate values, we find a deviation from the linear dependence, which is due to the increasingly facilitated vortex tailgating. We use these values of velocities and the frequency spectra to further discuss the potential for coherent radiation of vortices crossing the SN. In detail, considering the SN of width $w = 6\xi$, at $J = 0.348 J_{DP}$, the average velocity is v $\approx 0.03 \xi\tau_{GL}^{-1}$. The corresponding spectrum of modulations (figure \ref{fig:FFT6} - panel A) shows a number of contributions, the first five occurring at $\nu_{0}$ = 0.0021, $\nu_{1}$ = 0.0041, $\nu_{2}$ = 0.0062, $\nu_{3}$ = 0.0083, and $\nu_{4}$ = 0.0104 $\tau_{GL}^{-1}$, which are harmonics of the fundamental mode ($\nu_{0}$).
The period of the cycle of repeating vortex crossings in this case is T = 486 $\tau_{GL}$, while the wavelength is 14.6$\xi$ (obtained from $\lambda = v t$, using the value of vortex velocity in figure \ref{fig:vortex_velocity}). As the vortices move in a quasi-synchronous manner, the washboard frequency \cite{bulaevskii2006electromagnetic} is not applicable.\\
As the applied current is increased the vortices cross the SN in a more synchronized manner. In this case we can apply the relation for the washboard frequency to the values of the average vortex velocity $v$ and the frequency of first harmonic $\nu_0$ and obtain the value for the (virtual) lattice spacing $a$. Increasing $J$ from $0.366 J_{DP}$ to $0.655 J_{DP}$, the values of such lattice spacing decreases from 6.2$\xi$ to 2.8$\xi$, where values were extrapolated from the washboard frequency relation.
The combination of increasing Lorentz force and edge confining forces, cause the reduction of $a$ and raising of the vortex density during the dynamics. However, in the wider SN ($w = 12\xi$) we do not observe the same behaviour. In this case the value of $a$ remains constant ($\simeq 3.6 \xi$, obtained using the washboard frequency relation) as the current is increased in the dissipative state, until transitioning to asynchronous crossings for high $J$ values.  This suggests that in the wider SN the vortex-vortex repulsion within the lattice is more deterministic for the resulting lattice spacing $a$ than the interactions with confining edges during the vortex dynamics.
\begin{figure}[ht] 
	\centering
    	\includegraphics[width=\linewidth]{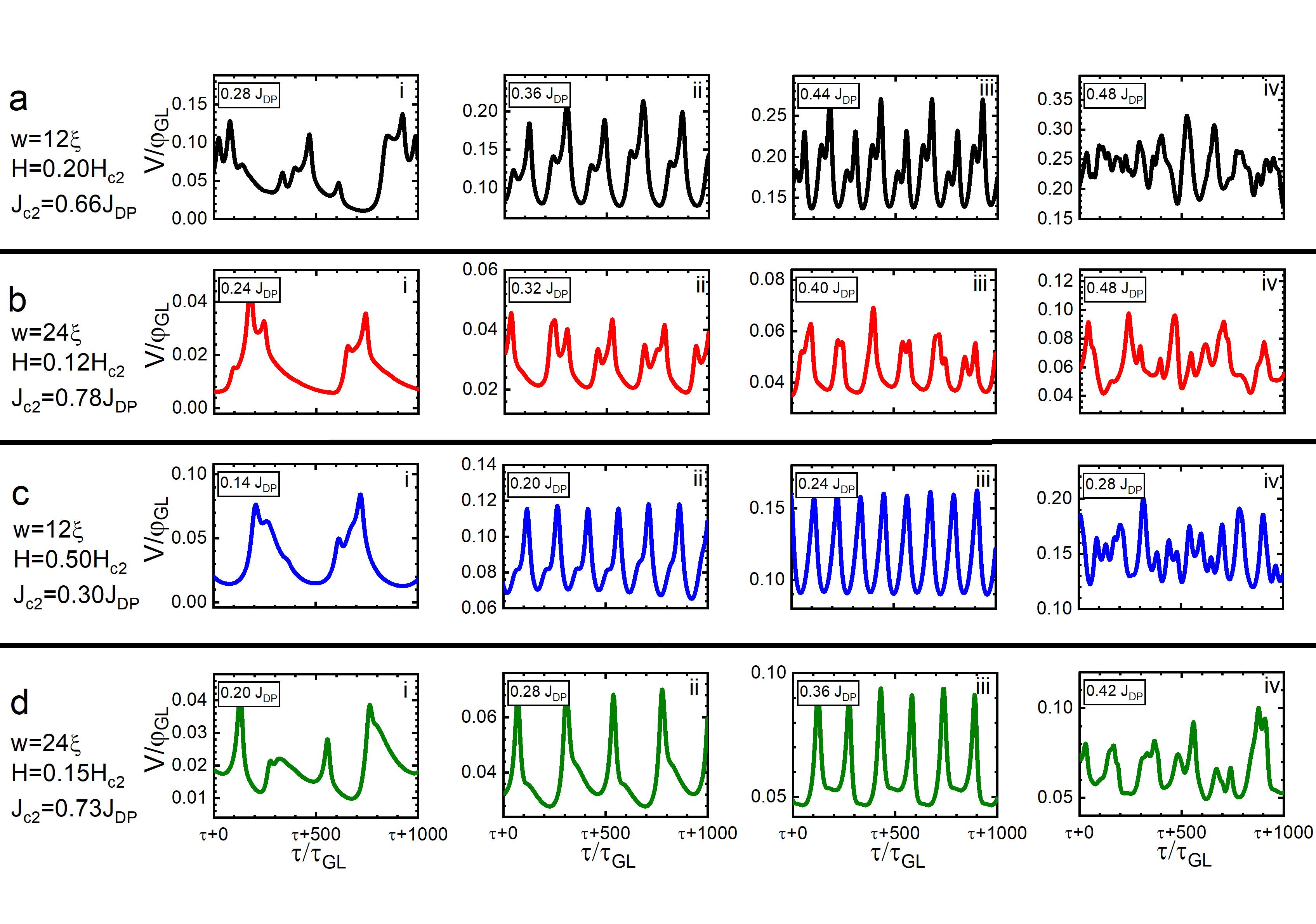} 
	\caption{Temporal evolution of the normalized voltage drop, at increasing (indicated) values of the applied current density for two SNs, of width 12$\xi$ and 24$\xi$. All plots exhibit voltage modulations caused by vortex crossing. Panels a.i-iv: $w = 12\xi$, $H = 0.20 H_{c2}$ (single row of vortices). Panels b.i-iv: $w = 24\xi$, $H = 0.12 H_{c2}$ (two vortex rows). Panels c.i-iv: $w = 12\xi$, $H = 0.50 H_{c2}$ (three rows). Panels d.i-iv: $w = 24\xi$, $H = 0.15 H_{c2}$ (three rows). The first panel in each row corresponds to the onset of the dissipative state; the second and third belong to the synchronous/quasi-synchronous regime; the fourth is the onset of the flux-flow instability regime.}
	\label{fig:all_modulations}
\end{figure}
The synchronisation of vortex crossing in a fixed lattice at large sourced currents was observed and discussed in ref. \cite{vodolazov2007rearrangement}, albeit without identifying an exact regime as such. To understand better the regime in which synchronous lattice crossings can occur, we performed a number of additional simulations. Figure \ref{fig:all_modulations} shows other four examples of the modulations in $V(t)$, corresponding to a SN with $w = 12\xi$ and $H = 0.2 H_{c2}$ (panels a.i-iv), which realises a single row of vortices in the static case (figure \ref{fig:phase_diagram}). At $J = 0.28 J_{DP}$ the SN becomes dissipative with asynchronous vortex-crossing behaviour (a.i), until at $J\simeq 0.36 J_{DP}$ quasi-synchronous crossings begin (a.ii). The latter continues (a.iii) until at $J = 0.48 J_{DP}$ the flux-flow instability sets in (a.iv), achieving the normal state at $J_{c2}=0.66 J_{DP}$. Panels b.i-iv of figure \ref{fig:all_modulations} show similar behaviour for a SN of width $w = 24\xi$ in applied magnetic field $H = 0.12 H_{c2}$ (realising two vortex rows), that will not transition to synchronised crossings as applied current is increased. However, after increasing the magnetic field applied to either SN, synchronous crossings will take place (panels c.i-iv and d.i-iv). For SN of $w$ = 12 and 24$\xi$ at $H$ = 0.50 and 0.15$H_{c2}$ respectively, the vortex configuration comprises three parallel rows. For $w = 12\xi$ ($w= 24 \xi$) synchronous crossings start at $J\simeq 0.20 J_{DP}$ ($J = 0.28 J_{DP}$) and continue up to the onset of flux-flow instability at $J = 0.28 J_{DP}$ ($J = 0.42 J_{DP}$). 

One concludes that the frequency of the radiation stemming from coherent vortex crossings can be tuned by $H$ or/and by $J$. The applied magnetic field $H$ changes the vortex density (affecting number of rows) and, hence, $a$; while transport current directly changes the vortex velocity. Both factors influence the behaviour of vortex crossings, which in turn affect the electromagnetic radiation emitted at frequencies $\nu =v/a$ \cite{vodolazov2007rearrangement}. For an insight into values expected in experiment, we consider the parameters measured for Nb thin films by Pinto \textit{et al.} \cite{pinto2018dimensional}. For example, in a Nb film of thickness $d$ = 20 nm, $\xi(0) \simeq$ 8.0 nm and $T_c = 8$K, providing a Ginzburg-Landau time of $\tau_{GL} \simeq 65$ fs, our results show an average velocity of vortices crossing the Nb SN of thickness 20nm and width 50-100 nm to be in the range 1-10 km/s, with first harmonic frequencies in the range 1-50 GHz. These values are similar to those reported by Dobrovolskiy \textit{et al.} \cite{dobrovolskiy2020ultra,dobrovolskiy2018microwave} and to those of Embon \textit{et al.} \cite{embon2017imaging}. 
A thin and narrow superconductor with high $T_c$ with a small value of $\tau_{GL}$ ($\simeq 1-10$fs), where faster vortex crossings could be realized and used as a terahertz radiation source. Such sources are highly sought for a variety of applications \cite{hafez2016intense}, including clinical \cite{son2019potential} and terahertz time-domain spectroscopy \cite{gowen2012terahertz}.

\section{Discussion}

This study has revealed a consistent theme, where the narrowest SNs exhibit stronger confinement forces. A vortex row phase diagram (figure \ref{fig:phase_diagram}) showed how the narrowest stripes support a lower number of vortex rows, and had a lower average vortex density (for given $H$) that deviated more from the theoretical value (figure \ref{fig:vortex_density}).
It was found a range of SN widths, 20-60$\xi$, where vortex rows would transition from a single to two-rows, and back to a single row (see inset figure \ref{fig:vortex_density}). This is due to an interplay of vortex interaction and the edge barrier strengths as the field is increased, below and above the range of width quoted the confinement forces are too strong or weak respectively. 
When studying the critical current as a function of magnetic field (figure \ref{fig:Jc_vs_H}), after a new row of vortices emerges $J_{c1}$ is at a minima, but increases as the field is increased due to stronger confinement forces acting a vortex pinning potential. It increases to a maxima before vortex density increases too high that the confinement forces at the edge overpower the vortex-vortex interaction. 
Vortices crossing the SN cause modulations in the voltage drop across the stripe (figures \ref{fig:modulations6xi}, \ref{fig:modulations12xi}, \ref{fig:all_modulations}), the same process also causes EM radiation to be emitted \cite{bulaevskii2006electromagnetic}. A single vortex crossing the stripe emits a photon with energy proportional to the inverse of the crossing time. More vortices crossing in synchronicity causes more power to be emitted for that frequency. Therefore, vortices moving in synchronous rows will provide greater radiation power. SNs were found to exhibit an evolution for how the vortex rows crossed the stripe as the applied current density was increased. Depending on the width and magnetic field, the SN would start dissipation with vortices crossing slowly quasi-synchronously (figure \ref{fig:modulations6xi},A) or with little repetition (figure \ref{fig:all_modulations},b). As $J$ increased the crossings progress to quasi-synchronous (figure \ref{fig:all_modulations},b) or synchronous (figure \ref{fig:modulations12xi},D), until proceeding to flux-flow instability. 
Not every example showed the evolution to synchronised vortex crossings, however, a regime for the occurrence of vortex rows crossing in a fixed lattice was observed for states with average vortex density $A \lesssim 81\xi^2$ (in figure \ref{fig:vortex_density}). This is also related to confinement, a large vortex density is required so the edge confining forces a can adequately act on the vortex rows and effectively lock them in a dynamic lattice. The more synchronised the vortex crossings, the few contributions to the spectrum of frequencies (figures \ref{fig:FFT6}, \ref{fig:FFT12}), which is beneficial for generation of EM radiation.

\bibliographystyle{unsrt}  
\bibliography{references}

\end{document}